\documentclass[a4paper]{article}

\usepackage{INTERSPEECH2019}
\usepackage{hyperref}

\title{Joint training framework for text-to-speech and voice conversion using multi-source Tacotron and WaveNet}
\name{Mingyang Zhang$^1$, Xin Wang$^2$, Fuming Fang$^2$,  Haizhou Li$^1$, Junichi Yamagishi$^2$}
\address{
  $^1$National University of Singapore, Singapore, 
  $^2$National Institute of Informatics, Tokyo, Japan
  }
\email{mingyang.zhang@u.nus.edu, \{wangxin,fang\}@nii.ac.jp, haizhou.li@nus.edu.sg, jyamagis@nii.ac.jp}

\begin{document}

\maketitle
\ninept
\begin{abstract}
We investigated the training of a shared model for both text-to-speech (TTS) and voice conversion (VC) tasks. We propose using an extended model architecture of Tacotron, that is a multi-source sequence-to-sequence model with a dual attention mechanism as the shared model for both the TTS and VC tasks. 
This model can accomplish these two different tasks respectively according to the type of input. An end-to-end speech synthesis task is conducted when the model is given text as the input while a sequence-to-sequence voice conversion task is conducted when it is given the speech of a source speaker as the input. 
Waveform signals are generated by using WaveNet, which is conditioned by using a predicted mel-spectrogram. We propose jointly training a shared model as a decoder for a target speaker that supports multiple sources. Listening experiments show that our proposed multi-source encoder-decoder model can efficiently achieve both the TTS and VC tasks.
\end{abstract}

\section{Introduction}
Text-to-speech (TTS) and voice conversion (VC) are two typical technologies for generating speech waveform. TTS is a technology that synthesizes natural-sounding human-like speech from text. 
The traditional approaches to TTS can be divided into two categories: a waveform concatenation approach and statistical parametric approach. Recently, along with developments on deep learning, neural-network-based end-to-end approaches have been proposed to achieve better performance, such as Tacotron \cite{wang2017tacotron}, DeepVoice/Clarinet \cite{ping2018deep}, and Char2Wav \cite{sotelo2017char2wav}.

Voice conversion is a technology that modifies the speech of a source speaker and makes their speech sound like that of another target speaker without changing the linguistic information. Many standard voice conversion approaches can be formulated as a regression problem of estimating a mapping function between the spectrum features of a source speaker and a target speaker. Gaussian mixed model (GMM) based approaches \cite{Toda2007, takamichi2015modulation, tanaka2017speaker} try to learn a linear transform from source to target features on the basis of the joint probability of these features by using a GMM. Dynamic kernel partial least squares regression (DKPLS) \cite{Helander2012} and neural network based approaches \cite{Chen2014, Nakashika2013, mohammadi2014voice, sun2015voice, zhang2018error, Sisman2018Inter} learn non-linear transforms. To make use of non-parallel source-target speech corpora, voice conversion based on the variational autoencoder (VAE) \cite{hsu2016voice} and cycle-consistent generative adversarial network \cite{8462342, hsu2017voice, kaneko2017parallel, 8639507} were proposed.

So far, even though various successful methods have been proposed for TTS and voice conversion, most of the systems can achieve only one task. For each problem, the network architecture is designed for the targeted task only and involves a long period of tuning specifically for the problem. This procedure needs to be repeated for different tasks, and this restrict the powerful effect of the neural network. 

However, we see that theoretical differences between VC and TTS are currently becoming much smaller than their original narrow definitions. To give a few examples, the recent advanced high-performance VC systems gain from the use of the phone posteriorgram (that is, a continuous phone representation) of inputted speech \cite{Liu2018}. There was also an attempt to use both the spectrum features and phone posteriorgram to further improve the performance of voice conversion \cite{8607053}. We can also see similar trends for TTS. The end-to-end TTS system sometimes also uses phone-embedding vectors as the input instead of letter inputs \cite{ping2018deep,li2018close}. There was also an attempt to use a reference audio signal as the additional input for Tacotron to transfer the prosody of the reference audio into synthetic speech via a reference encoder \cite{ReferenceEncoder}\footnote{There is also a task called prosody conversion in the VC field \cite{Sisman2018}.}.  
Given the above trends, we strongly believe that we can construct one model shared for both the TTS and VC tasks. This idea is reasonable since any improvement of the proposed model is expected to improve both tasks and we can also utilize speech databases for both tasks at the same time. This will also give us a good opportunity to deeply re-consider fundamental differences between VC and TTS.   

To achieve this goal, we propose a multi-source sequence-to-sequence model with dual attention mechanism. We assume that both TTS and voice conversion can be divided into two parts: an input encoder and an acoustic decoder. The difference between the two tasks is that the input of TTS is text characters while that of voice conversion is acoustic features. The model can be thought of as an encoder-decoder model that supports multiple encoders. The role of multiple encoder networks is the frond-end processing of each type of input data and the role of a decoder network is to predict acoustic features required for waveform generation. Inspired by the success of end-to-end TTS models, we adopt architectures similar to Tacotron for the encoders and decoder. More specifically, we have two encoders that encode different inputs and a shared decoder that predicts the acoustic features, followed by the generation of high-quality waveform signals based on WaveNet, a generative model for raw audio waveforms \cite{van2016wavenet}.
The other contributions of our work are as follows. First, to achieve better TTS performance with a small amount of training data, we adapt a pre-trained TTS model to a target speaker. Second, for voice conversion, we train a many-to-one conversion to increase the size of training data while restricting the use of parallel data.

The rest of the paper is arranged as follows. Section 2 introduces the end-to-end speech synthesis model Tacotron. In Section 3, we describe our proposed joint training model for TTS and voice conversion. Section 4 presents experiments and the results of a large-scale listening test. We conclude our paper and discuss future work in Section 5.

\section{Tacotron}
\begin{figure}
	\centering
    \includegraphics[width=65mm]{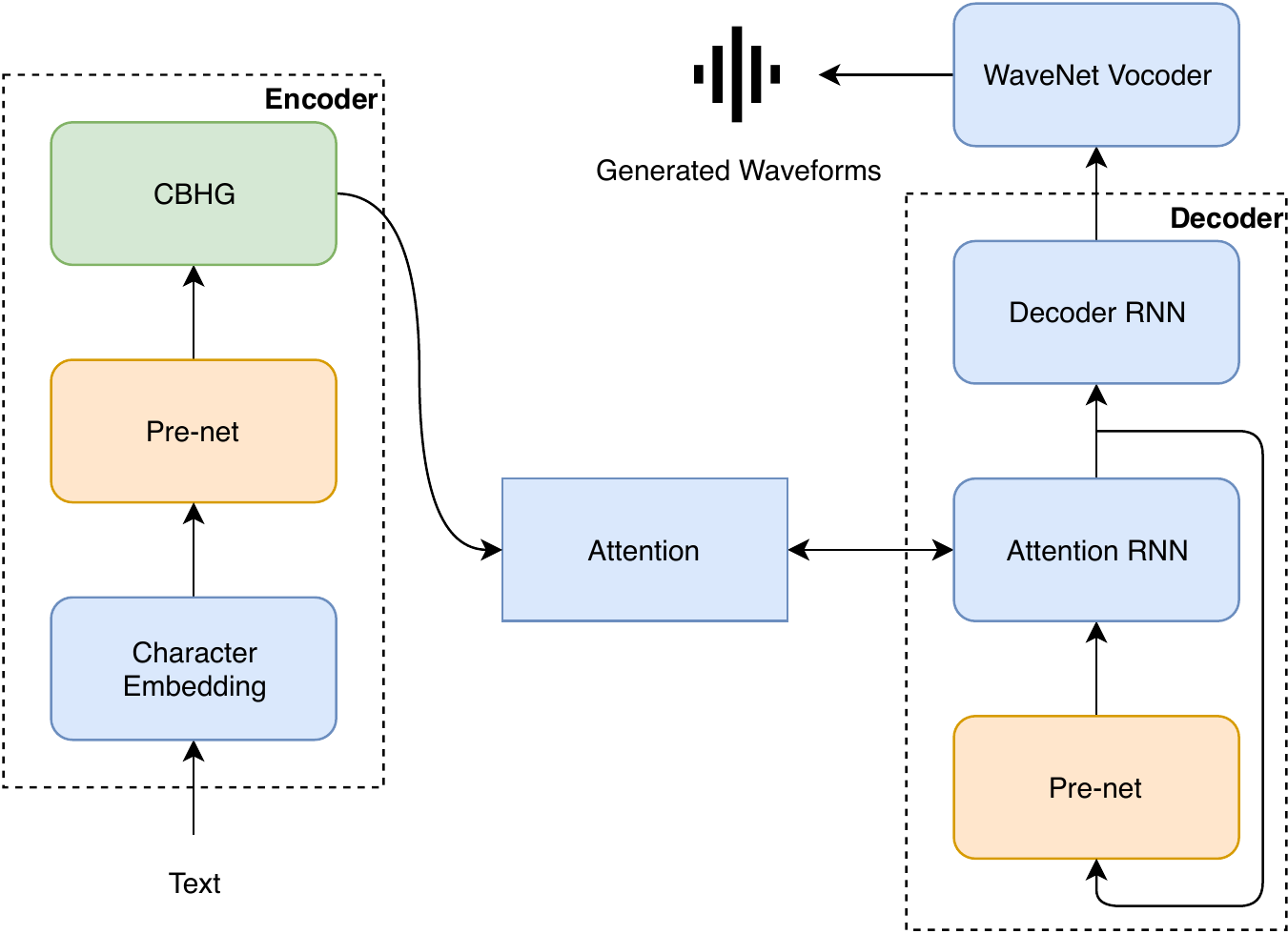}
    \caption{Block diagram of Tacotron system with WaveNet vocoder}
    \vspace{-3mm}
    \label{fig:tacotron}
    \vspace{-3mm}
\end{figure}

Tacotron is an end-to-end text-to-speech (TTS) system that synthesizes speech directly from characters \cite{wang2017tacotron}. The architecture of Tacotron is a sequence-to-sequence model with an attention mechanism \cite{bahdanau2014neural}. Figure \ref{fig:tacotron} illustrates a diagram of Tacotron, which includes an encoder that maps an input text sequence to a fixed-dimensional state vector and an attention-based decoder that predicts a mel spectrogram. The WaveNet vocoder is conditioned on the predicted mel spectrogram to generate speech waveforms.

\vspace{-3mm}
\subsection{Architecture}
\vspace{-1mm}

The input of the encoder is a sequence of characters, where each character is encoded as a one-hot vector and embedded into a continuous vector. These embedded vectors are then processed by a pre-net that contains a bottleneck layer with dropout. The outputs of the pre-net are used as the input to a CBHG module to produce the final results of the encoder. The CBHG module consists of a convolutional bank, highway network and bidirectional gated recurrent unit (GRU).

The decoder is an auto-regressive (AR) recurrent neural network (RNN) that predicts a sequence of mel spectrum vectors from the encoder outputs. A Bahdanau attention mechanism is used to summarize the encoder output as a fixed-length context vector. The concatenation of the context vector and the output of an attention RNN are used as the input to the decoder RNN. The target feature of the decoder is a mel spectrogram. At run-time, the predicted mel spectrum vector of the previous time step is fed to the pre-net to generate the input of the attention RNN. During training, the ground truth mel spectrum of the corresponding frame is fed to the pre-net.

\vspace{-3mm}
\subsection{Adaptation of Pre-trained Tacotron Model}
\vspace{-1mm}
Training a Tacotron-based TTS system from scratch normally requires a large amount of data, for example, usually at least a few hours of speech \cite{wang2017tacotron}, even though there is the know-how to reduce the amount of required data, such as decoder pre-training \cite{chung2018semi}. In this study, we assume that the amount of speech data available for a target speaker is as small as the usual VC task and is less than one hour. 

In this situation, it is reasonable for us to conduct speaker adaptation by using a Tacotron model that is well- and pre-trained with a target speaker in order to compensate for the data size restriction. 
Such adaptation of a pre-trained speech synthesis model usually involves two steps: training a high-performance model by using large-scale speech corpora and using the well-trained model as the initial seed model and fine-tuning with a small amount of training data from the target speaker. In our case, we pre-trained a Tacotron by using the LJ Speech database and adapted it to a different female speaker included in the CMU ARCTIC database.

\section{Joint Training of TTS \& VC}
\begin{figure*}
    \centering
    \includegraphics[width=100mm]{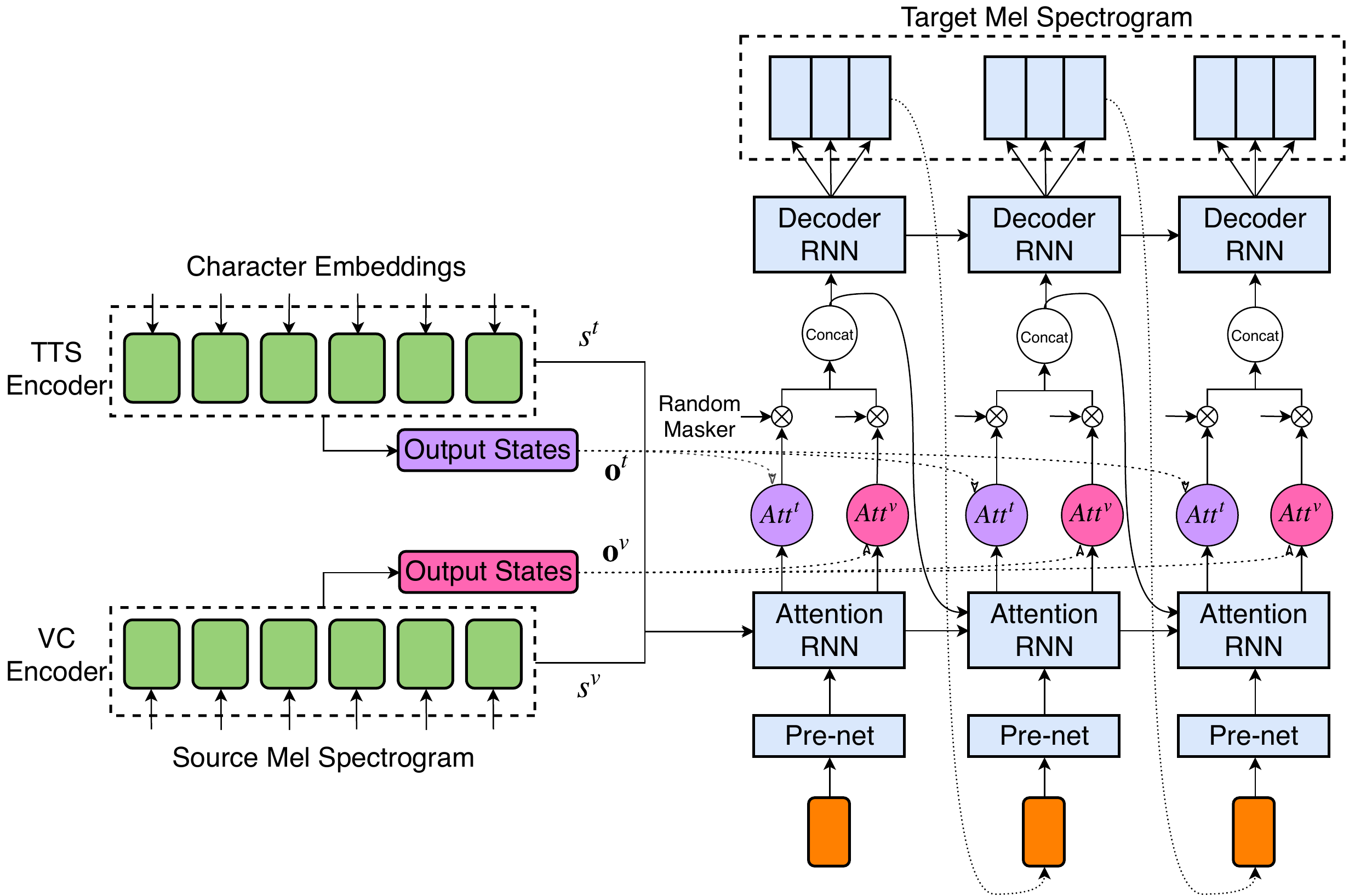}
    \vspace{-3mm}
    \caption{Model architecture of our proposed multi-source sequence-to-sequence model for training TTS \& VC simultaneously. Random maskers are applied to all decoder steps.}
    \label{fig:model}
    \vspace{-5mm}
\end{figure*}

Our proposed multi-source Tacotron model is illustrated in Figure \ref{fig:model}. It consists of a TTS input encoder, a VC input encoder, and a dual attention mechanism-based acoustic decoder followed by a WaveNet vocoder.

\vspace{-3mm}
\subsection{Encoders}
\vspace{-1mm}

Both TTS input and VC input encoders have the same architecture, which includes a pre-net and a CBHG network. Each encoder transforms the corresponding input sequences into a fixed dimension state vector and a set of encoder output vectors:
\begin{align}
    s^{t},\mathbf{o^{t}}&=Encoder(\mathbf{x^{t}})\\
    s^{v},\mathbf{o^{v}}&=Encoder(\mathbf{x^{v}}),
\end{align}
where $\mathbf{x^{t}}=\{x_{1}^{t}, x_{2}^{t}, ..., x_{M}^{t}\}$ denotes the character embedding sequence of length $M$ that is the input to the TTS encoder, $\mathbf{x^{v}}=\{x_{1}^{v}, x_{2}^{v}, ..., x_{N}^{v}\}$ denotes the source mel spectrogram of $N$ frames that is the input to the VC encoder, $s^{t}$ and $\mathbf{o^{t}}$ respectively represent the state vector and output sequence of the TTS encoder, and $s^{v}$ and $\mathbf{o^{v}}$ respectively represent the state vector and output sequence of the VC encoder. 

\vspace{-3mm}
\subsection{Dual Attention-based Decoder}
\vspace{-1mm}
The decoder of our model is based on the decoder of Tacotron, and it consists of a pre-net, an attention RNN layer, and a decoder RNN layer. Since a character embedding sequence and mel spectrogram have different time scales and we have to cope with the asynchronous input sequences, we use a dual attention mechanism. Two independent attention mechanisms $Att^{t}$ and $Att^{v}$ are used for transforming the outputs of the TTS and VC input encoders into context vectors, respectively. 

At every output step $k$, the attention RNN produces the state $h_{k}^{a}$ and the output vector $o_{k}^{a}$, given the previous step state $h_{k-1}^{a}$, context vectors $c_{k-1}^{t}$ and $c_{k-1}^{v}$, and pre-net output $o_{k}^{p}$. Attention vector are generated from the output state and the encoder output vectors $\mathbf{o^{t}}$ and $\mathbf{o^{v}}$, and they are then combined with the encoder outputs to produce the context vectors $c_{k}^{t}$ and $c_{k}^{v}$. The two context vectors and the attention RNN output are concatenated as the input of the decoder RNN. We use a fully-connected layer to predict the $r$ frames of the mel spectrogram of the target speech data from the decoder output $o_{k}^{d}$, represented as $\{\hat{y}_{(k-1)r+1}, ..., \hat{y}_{kr}\}$. The ground truth of the last predicted frame is fed to the pre-net to produce the next step output. At the first output step, the concatenation of the final encoder states $s^{v}$ and $s^{t}$ is used as the input of the attention RNN instead of the previous step state. The procedures of the dual attention-based decoder are summarized as follows.
\begin{align}
    o_{k}^{p} &= PreNet(y_{(k-1)r})\\
    h_{k}^{a}, o_{k}^{a} &= AttRNN(h_{k-1}^{a}, c_{k-1}^{t}, c_{k-1}^{v}, o_{k}^{p})\\
    c_{k}^{t} &= \mathbf{o^{t}} \cdot Att^{t}(h_{k}^{a}, \mathbf{o^{t}})\\
    c_{k}^{v} &= \mathbf{o^{v}} \cdot Att^{v}(h_{k}^{a}, \mathbf{o^{v}})\\
    o_{k}^{d} &= DecoderRNN(h_{k-1}^{d}, o_{k}^{a}, c_{k}^{t}, c_{k}^{v})\\
    \{\hat{y}_{(k-1)r+i}\} &= fc(o_{k}^{d})\qquad(i=1, ..., r)
\end{align}

\subsection{Random Selection of Input Encoders}
\vspace{-1mm}
Networks with multi-source inputs can often be dominated by one of the inputs \cite{feichtenhofer2016convolutional}. In our proposed framework, since mapping from a source mel spectrogram to target mel spectrogram is much easier than mapping from a character embedding to a target mel spectrogram, the model will be dominated by the mel spectrogram input. To alleviate this problem, one of the following input types is randomly chosen during training: character embedding only, source mel spectrogram only, or both of the inputs to ensure that we only give specific input information to the decoder. To achieve this, we introduce a random masker for indicating which input to use during the training. Using the masker, we set the context vector that belongs to the unused input types to zero. 

\vspace{-3mm}
\subsection{Training Procedures of Joint Model}
\vspace{-1mm}
Even with this strategy, when the model is trained from scratch, we still encounter the same problem, meaning that only mapping from spectrogram to spectrogram works. To prevent this from happening, we conduct the following step-by-step training. We first train the two tasks' models separately and then use the encoder from each trained model to initialize the encoders of the multi-source model. Here, the TTS stand-alone model is adapted from a pre-trained TTS as described in Section 2. The VC stand-alone model is a many-to-one VC trained with parallel utterances from multiple source speakers and a target speaker. 

Then, we jointly train the multi-source model with two inputs by using the dual attention mechanism. This mechanism allows the model to extract information from both character embedding and mel spectrogram inputs, even when one of them is absent, or the two of them are not time aligned.

\vspace{-1mm}
\subsection{Generation Stage}
\vspace{-1mm}
Different from the teacher-forcing training method, 
the model at run-time does not have a ground truth to feed to the pre-net in the decoder.
Therefore, we use the last frame of the $r$ previously predicted frames as the input of the pre-net. Then, equation (3) now becomes as follows.
\begin{equation}
    o_{k}^{p} = PreNet(\hat{y}_{(k-1)r})
\end{equation}
Given the different kinds of input to our proposed framework, we can choose which task should be achieved by setting the masker. If we use only the character embedding input, the system becomes a TTS model. If we use only the source mel spectrogram input, the system becomes a VC model. If we use both of the inputs, we can see this as a hybrid model of TTS and VC. We will compare the performance of these models in Section 4.

\section{Experiment}

\subsection{Datasets}
\vspace{-2mm}
For the training of the pre-trained TTS model, we used the LJ Speech database, which consists of 13,100 short audio clips of a single speaker reading passages from 7 non-fiction books and has a total length of approximately 24 hours \cite{ljspeech17}. For the adaptation of the pre-trained TTS model, we used 500 utterances of a female speaker, SLT, from the CMU ARCTIC database \cite{Kominek03cmuarctic}.

For the training of the many-to-one VC model, we conducted a mapping from male speakers to a female speaker. The source speakers were two male speakers, BDL and CLB, from the CMU ARCTIC database, and the target speaker was SLT. We used exactly the same 500 utterances from SLT that were used to adapt TTS as the target speech data. The source speech data was the parallel data of the target speech from the two male speakers. This increased the training data size, and we had a total of 1,000 utterances.
\vspace{-2mm}
\subsection{Experimental Setup}
\vspace{-2mm}
We extracted 80-band mel spectrograms from all of the speech data as the source and target spectrum features. The frame length was 50ms and the frame shift was 12.5ms. We converted the input text characters into 256-dimension character embedding vectors to form the inputs of the text input encoder. We set the reduction factor parameter $r=2$ to predict two frames of mel spectrogram in one step.

To measure the performance of our proposed framework, we evaluated our model and compared it with the following systems.
\begin{itemize}
    \item \textbf{TTS}: Stand-alone model of adapted TTS system
    \item \textbf{VC}: Stand-alone many-to-one VC model using same source speakers and target speaker
    \item \textbf{Hybrid TTS}: Proposed model with only text input
    \item \textbf{Hybrid VC}: Proposed model with only source speaker's speech input
    \item \textbf{Hybrid TTS \& VC}: Proposed model with both text and source speaker's speech inputs
\end{itemize}
In addition to the above synthesis/conversion systems, we further added two systems for calibration: one for two source speakers and one for the target speaker. Therefore, we had in total seven systems for comparison.
We randomly selected 132 sentences or 132 pieces of source speech for synthesizing or converting the inputs\footnote{Audio samples of synthetic and convert speech are available at  \href{https://nii-yamagishilab.github.io/hybrid-TTS-VC/}{\underline{here}}.}. 
Note that 66 source speech waveforms came from one source speaker, BDL, while the rest were from the other source speaker, CLB.

We trained our model with a batch size of 32 by using the Adam optimizer with $\beta_{1}=0.9, \beta_{2}=0.999$, and the initial learning rate was 0.002 with the Noam decay scheme \cite{vaswani2017attention}.
In the model architecture, the pre-net was a dense layer with a dropout of 0.5, and the output dimension was 128. In the CBHG module, the 1-D convolutional bank was 16 sets of 1-D convolutional filters with 128 output channels with ReLU activation. The convolutional outputs were max pooled with a stride of 1 and a width of 2. Then, we passed the sequences to a convolution layer with width of 3 and 128 output channels with ReLU activation, followed by a convolution layer with width of 3 and 128 output channels. The highway network consisted of 4 layers of fully-connected layers with a 128 output dimension and ReLU activation. The bidirectional GRU RNN layer had 128 cells. The attention RNN was a 1-layer GRU with 256 cells. The decoder RNN was a 2-layer residual GRU with 256 cells.

The five experimental systems used the same WaveNet-vocoder to generate the speech waveforms with a sampling rate of 16k Hz and 10-bits $\mu$-law quantization. This WaveNet-vocoder processed an input mel spectrogram by using a bi-directional LSTM recurrent layer and a one-dimensional convolution layer with a window size of 3, after which it upsampled the processed features to 16k Hz. Both the LSTM and convolution layers had a layer size of 64. The WaveNet-vocoder contained 30 dilated convolution layers, and the $k$-th layer had a dilation size of $2^{\mathrm{mod}(k-1,10)}$, where $\mathrm{mod}(\cdot)$ was the modulo operation. The output of the dilated convolution layers and that of the skip channel had 64 and 256 dimensions, respectively. The network was trained by using Adam with a learning rate of 0.0003.

\subsection{Listening Experiments}

The quality of the speech samples and its similarity to the target speaker were evaluated in a 1-to-5 Likert mean opinion score (MOS) test.
This evaluation was carried out with a web-based interface on a crowdsourcing platform. 
On each web page, we presented two questions, one for speech quality and the other for speaker similarity.
The evaluators were required to fully listen to one sample before evaluating it. Each evaluator could evaluate at most 210 speech samples. 
In total, 97 evaluators participated in the test and  produced a total of 18480 MOS scores. Accordingly, each speech sample received 10 quality and 10 similarity scores.

\begin{figure}
    \centering
    \includegraphics[width=80mm]{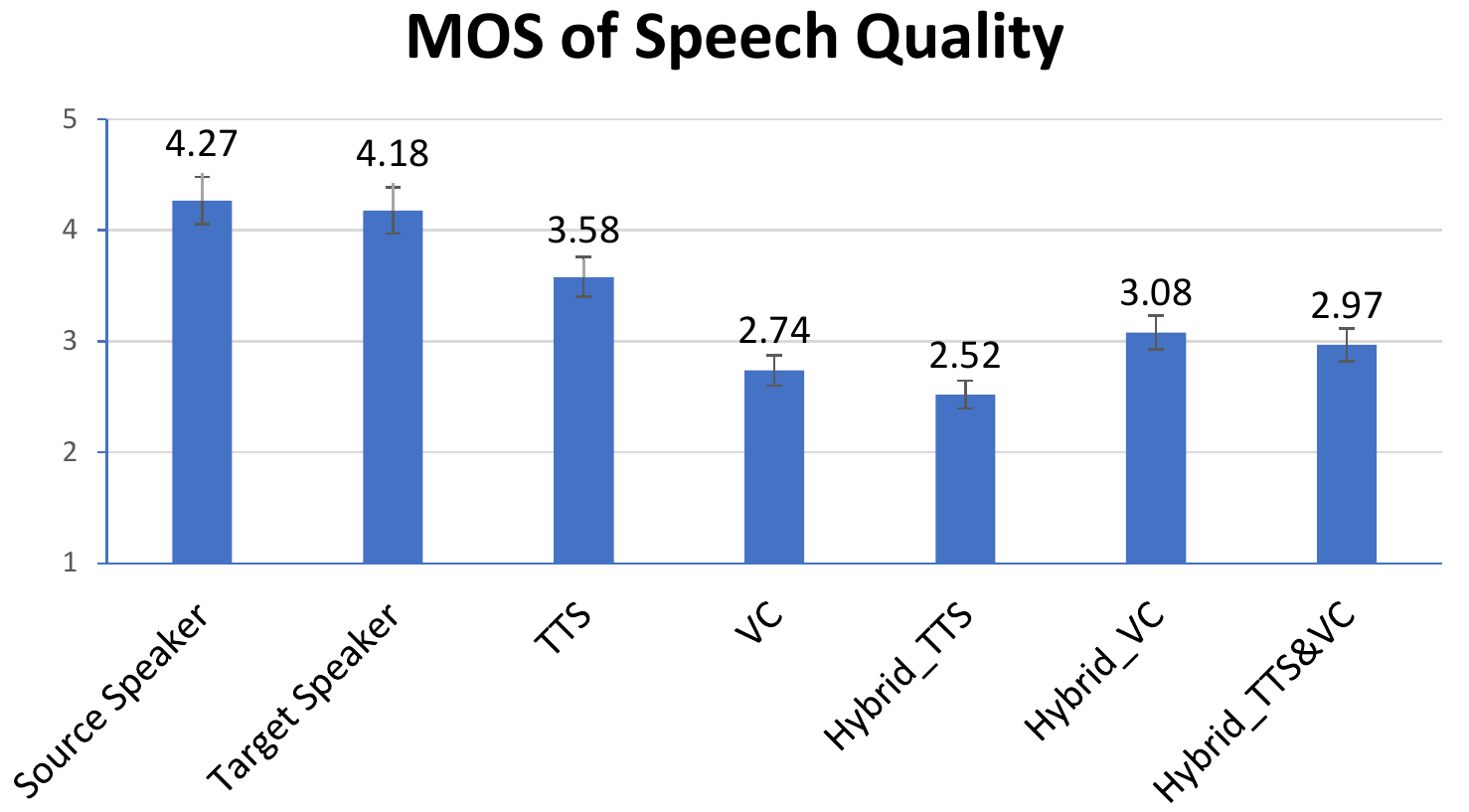}
    \vspace{-5mm}
    \caption{MOS results with 95\% confidence intervals for speech quality of different models}
    \label{fig:MOS_quality}
    \vspace{3mm}
    \includegraphics[width=80mm]{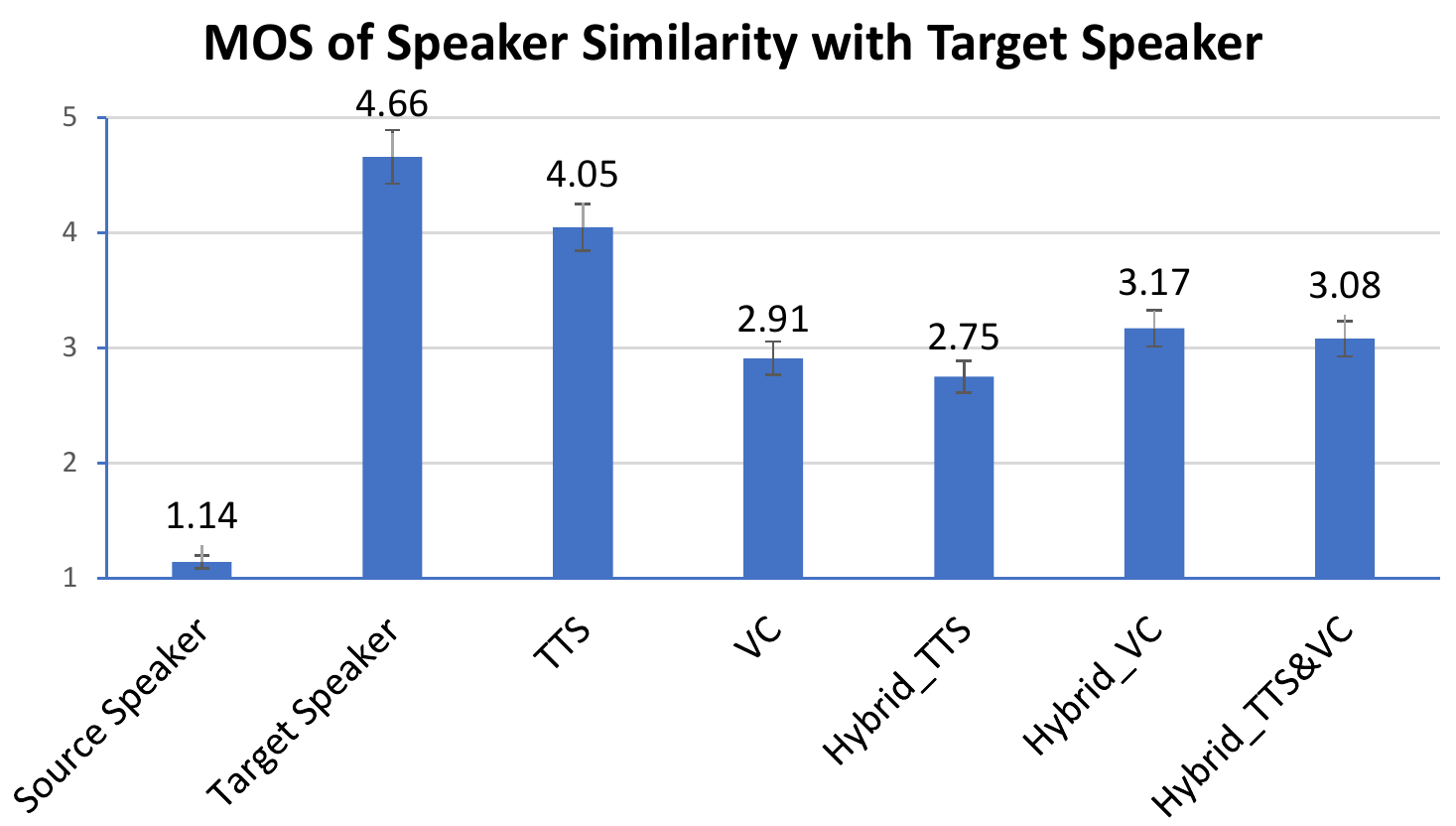}
    \vspace{-5mm}
    \caption{MOS results with 95\% confidence intervals for speaker similarity of different models}
    \label{fig:MOS_similarity}
    \vspace{-3mm}
\end{figure}

The results for speech quality and speaker similarity are shown in Figure~\ref{fig:MOS_quality} and Figure~\ref{fig:MOS_similarity}, respectively. It was observed that our proposed model worked for both the TTS and VC tasks. We can see that the hybrid VC system outperformed the VC stand-alone system in terms of both speech quality and speaker similarity. This indicates that our proposed model improved the performance of VC. However, the MOS results for the hybrid TTS system were worse than those for the TTS stand-alone system. We can hypothesize several reasons for this. First, the current multi-source model might still be over-fitting to the VC task. Second, it might not have sufficient parameters for doing both the TTS and VC tasks. We may need to increase the number of parameters especially for the TTS task. Third, random selection may not be the best strategy for the maskers of the input encoders. Better scheduling of the maskers needs to be investigated.

\section{Conclusion and Future Work}

In this paper, we proposed a joint model for both the TTS and VC tasks. The architecture of our model is based on Tacotron. Given text characters as input, the model conducts end-to-end speech synthesis. Given the spectrogram of a source speaker, the model conducts sequence-to-sequence voice conversion. The experimental results showed that our proposed model achieved both TTS and VC tasks and improved the performance of VC compared with the stand-alone model. Our future work will be to investigate a better method for the maskers of the input encoders and a more appropriate training algorithm.

\section{Acknowledgements} 
This research was carried out while the first author was at NII, Japan in 2018 using NII International Internship Program. 
This work was partially supported by JST CREST Grant Number JPMJCR18A6 (VoicePersonae project), Japan and by MEXT KAKENHI Grant Numbers (16H06302, 17H04687, 18H04120, 18H04112, 18KT0051), Japan.

\bibliographystyle{IEEEtran}

\bibliography{main}


\end{document}